\documentclass[conference,a4paper]{IEEEtran}
\usepackage{xcolor}
\usepackage{balance}

\ifCLASSINFOpdf
  \usepackage[pdftex]{graphicx}
\else
  \usepackage[dvips]{graphicx}
\fi
\ifCLASSOPTIONcompsoc
 \usepackage[caption=false,font=normalsize,labelfont=sf,textfont=sf]{subfig}
\else
 \usepackage[caption=false,font=footnotesize]{subfig}
\fi
\usepackage{dblfloatfix}

\usepackage{cite}
\usepackage{amsmath} 

\begin{document}
%
\title{A 100-GHz CMOS-Compatible RIS-on-Chip Based on Phase-Delay Lines for 6G Applications}

\author{\IEEEauthorblockN{
Xiarui Su,   
Xihui Teng,   
Yiyang Yu,
Yiming Yang,
Atif Shamim    
}                                     
\IEEEauthorblockA{
Electrical Engineering Department, King Abdullah University of Science and Technology (KAUST), Thuwal, Saudi Arabia, 
\\Email address: xiarui.su@kaust.edu.sa}

}



\maketitle

\begin{abstract}
On-chip reconfigurable intelligent surfaces (RIS) are expected to play a vital role in future 6G communication systems. This work proposed a CMOS-compatible on-chip RIS capable of achieving beam steering for the first time. The proposed unit cell design is a combination of a slot, a phase-delay line with VO$_2$, and a ground. Under the two states of the VO$_2$, the unit cell has a 180$^\circ$ phase difference at the center frequency, while maintaining reflection magnitudes better than -1.2 dB. Moreover, a 60×60 RIS array based on the present novel unit is designed, demonstrating the beam-steering capability. Finally, to validate the design concept, a prototype is fabricated, and the detailed fabrication process is presented. The measurement result demonstrates a 27.1 dB enhancement between ON and OFF states. The proposed RIS has the advantages of low loss, CMOS-compatibility, providing a foundation for future 6G applications.
\end{abstract}

\vskip0.5\baselineskip
\begin{IEEEkeywords}
CMOS-compatible design, 6G applications, phase-delay line, reconfigurable intelligent surface (RIS).
\end{IEEEkeywords}

%

\section{Introduction}
\begin{figure}[htp]
\centering
\includegraphics[width=\columnwidth]{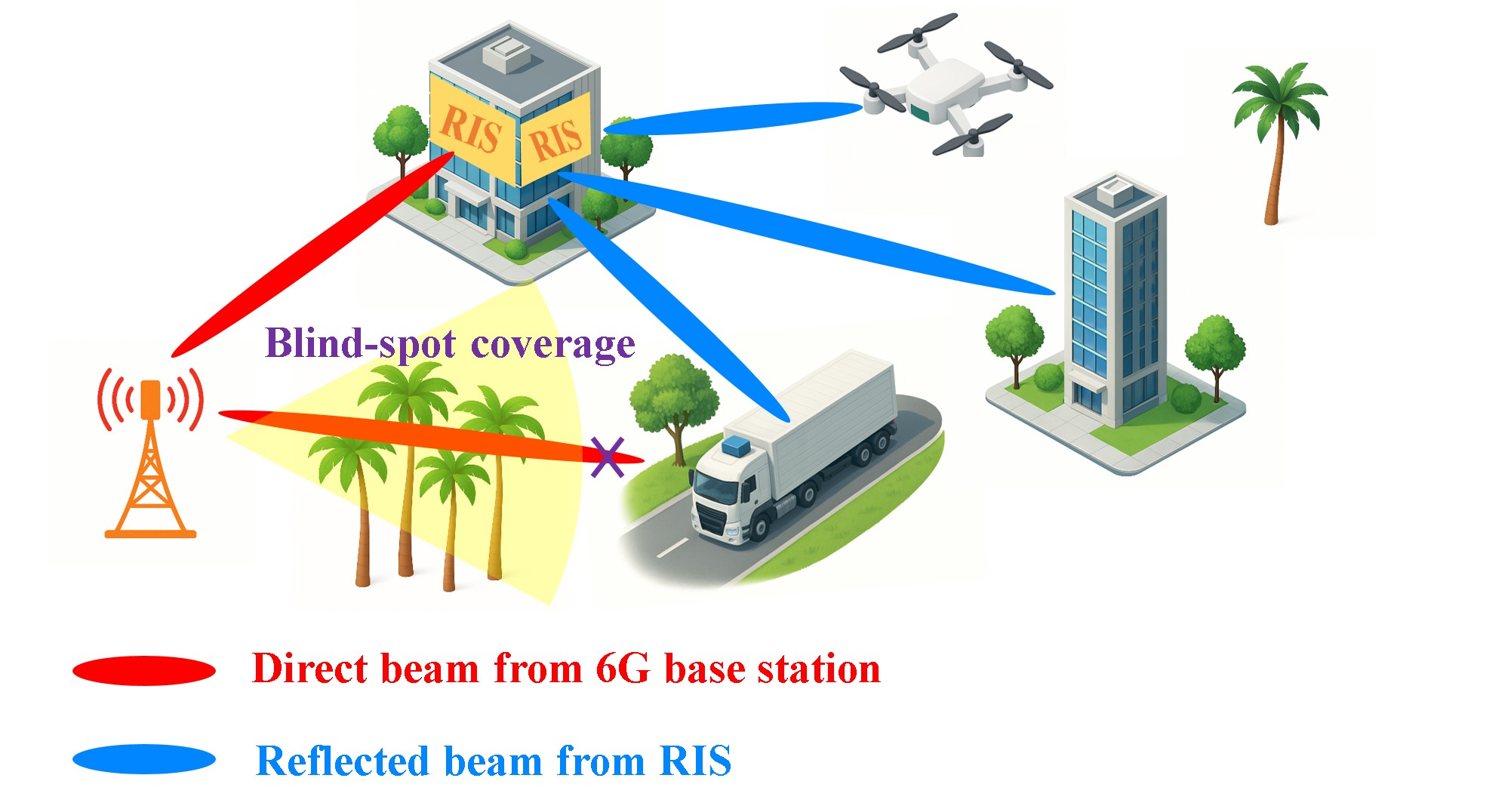}
\caption{Conceptual illustration of practical RIS implementations.
}
\label{fig1}
\end{figure}

Reconfigurable intelligent surfaces (RIS) have emerged as a key technology for 6G applications, by making the electromagnetic environment programmable. By electronically tuning its reflective or transmissive response, RIS can shape the propagation of radio waves to improve signal quality and coverage. As illustrated in Fig.~\ref{fig1}, RIS can be strategically deployed on buildings or infrastructure to assist base stations in overcoming signal blockages and providing coverage to aerial and ground users. With the advancement of 6G communications, reconfigurable intelligent surfaces (RIS) are expected to support higher data rates by operating at higher frequencies, maintaining low insertion loss, as well as having the compatibility with standard semiconductor fabrication processes.

In recent years, most reconfigurable intelligent surface (RIS) designs have realized phase control at relatively low frequencies by loading diodes, typically implemented through printed circuit board (PCB) fabrication techniques \cite{sheng2025,machado2024}. These designs demonstrate excellent performance, including superior beam-scanning capability and enhanced coverage. However, employing a large number of diodes considerably increases the overall cost and necessitates additional, complex control circuitry. Additionally, based on the PCB technology, the processing tolerances make it hard to guarantee the complicated structures at high frequencies. Recently, efforts have been made to extend RIS operation toward higher frequencies using materials such as liquid crystals and phase-change media \cite{Neuder2025,cui2023,xu2025}. Although these approaches enable tunable phase control at sub-terahertz or terahertz frequencies, they often suffer from high insertion loss, slow response, and limited compatibility with standard CMOS processes. Therefore, it remains a challenge to design RIS that can simultaneously achieve low loss in high frequency and good compatibility with semiconductor fabrication processes.

To address these challenges, this paper proposes a CMOS-compatible on-chip RIS based on phase-delay lines for 6G applications. A phase-delay line loaded with VO$_2$ is designed on a silicon substrate, achieving binary phase control with low loss. Furthermore, the whole structure is fully compatible with standard microfabrication processes. For further demonstration, a 60×60 RIS array with the proposed techniques is designed and fabricated.
\section{RIS Design}

\subsection{Element Analysis}
\subsubsection{Design of element}
\begin{figure}[t]
\centering
\includegraphics[width=\columnwidth]{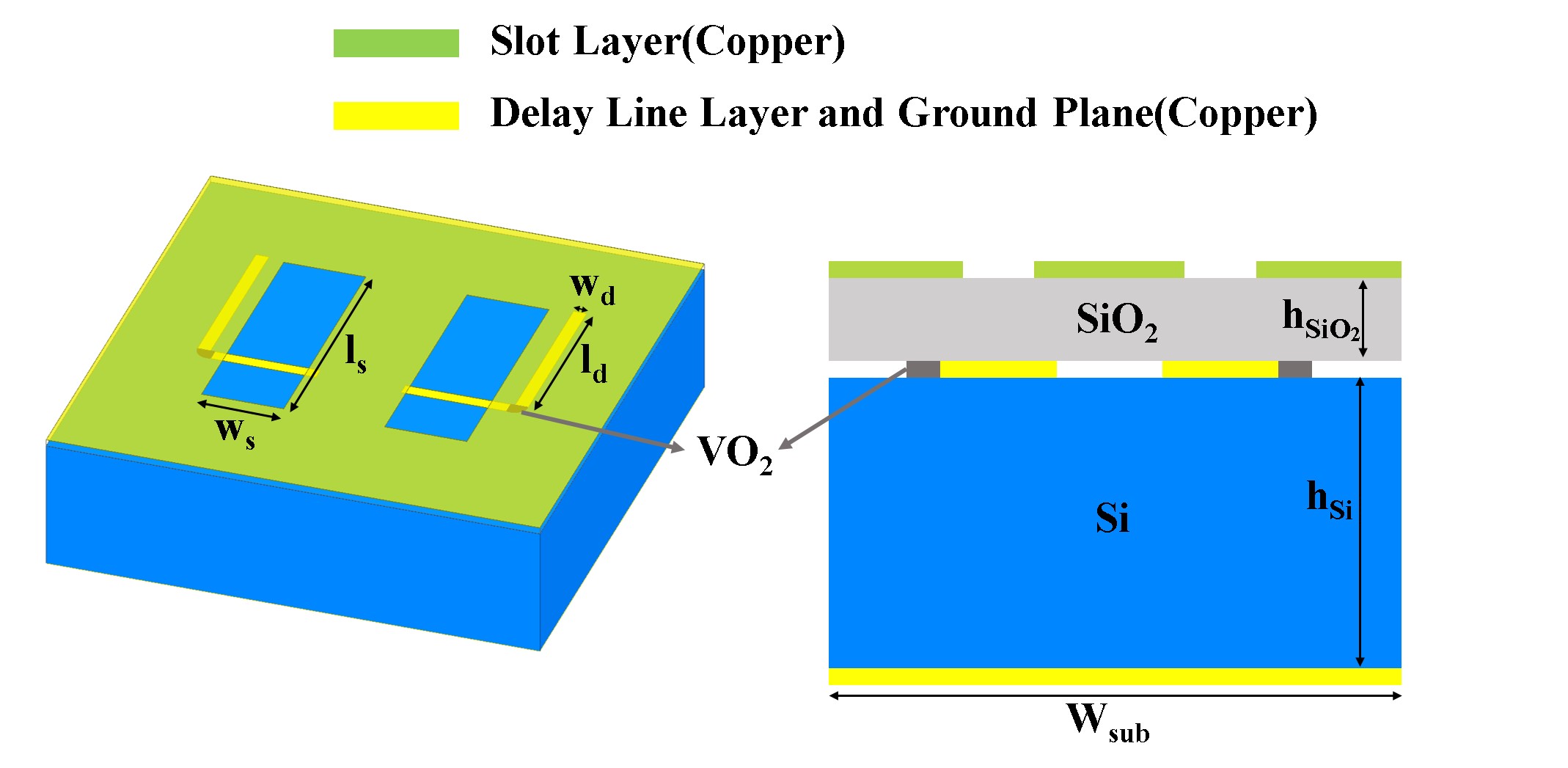}
\caption{Schematic illustration of the proposed RIS-on-chip unit cell showing its 3D structure and cross-sectional view.}
\label{fig2}
\end{figure}

The stackup for the unit cell is shown in Fig.~\ref{fig2}. The unit metallic structure is composed of a slot, a phase-delay line loaded with VO$_2$, a layer of silicon dioxide (SiO$_2$), and a layer of silicon. The slot layer on the top acts as a frequency-selective structure. Its width determines the center frequency where the phase transition occurs. The slot also serves as a coupling window, transferring the incident wave from the air side to the underlying delay line through capacitive near-field coupling.
Beneath the slot, the SiO$_2$ layer provides both electrical and dielectric isolation. It prevents direct conduction between the slot and the metal layer below, while reducing coupling to the high-permittivity silicon substrate and minimizing dielectric loss.
The delay-line layer guides surface current along paths of different lengths, generating two reflection-phase states and realizing binary phase control.
The VO$_2$ section bridges the metallic lines and switches between conducting and insulating states, changing the effective current path length and producing an approximately 180° phase shift.
The ground plane at the bottom serves as a reflective surface, while the silicon substrate offers mechanical support and full compatibility with CMOS and standard microfabrication processes.

The parameters utilized are summarized in Table~\ref{table1}. The delay-line length $l_d$ is designed to provide a 180$^\circ$ phase difference between the two switching states at the target frequency. The line width can also be adjusted to optimize the reflection response, as impedance matching to 50~$\Omega$ is not required for reflectarray operation.

 \begin{table}
\renewcommand{\arraystretch}{1.3}
\caption{Parameters of the proposed unit cell.}
\label{table1}
\centering
\begin{tabular}{|c|c|}
\hline
Parameter & Value\\
\hline
$W_{sub}$ & 1.125 mm\\
\hline
$h_{si}$ & 300 um\\
\hline
$h_{sio2}$ & 15 um\\
\hline
$w_s$ & 187.5 um\\
\hline
$l_s$ & 562.5 um\\
\hline
$w_d$ & 30 um\\
\hline
$l_d$ & 400 um\\
\hline
\end{tabular}
\end{table}

\subsubsection{Simulations}
\begin{figure}[htp]
\centering
\includegraphics[width=\columnwidth]{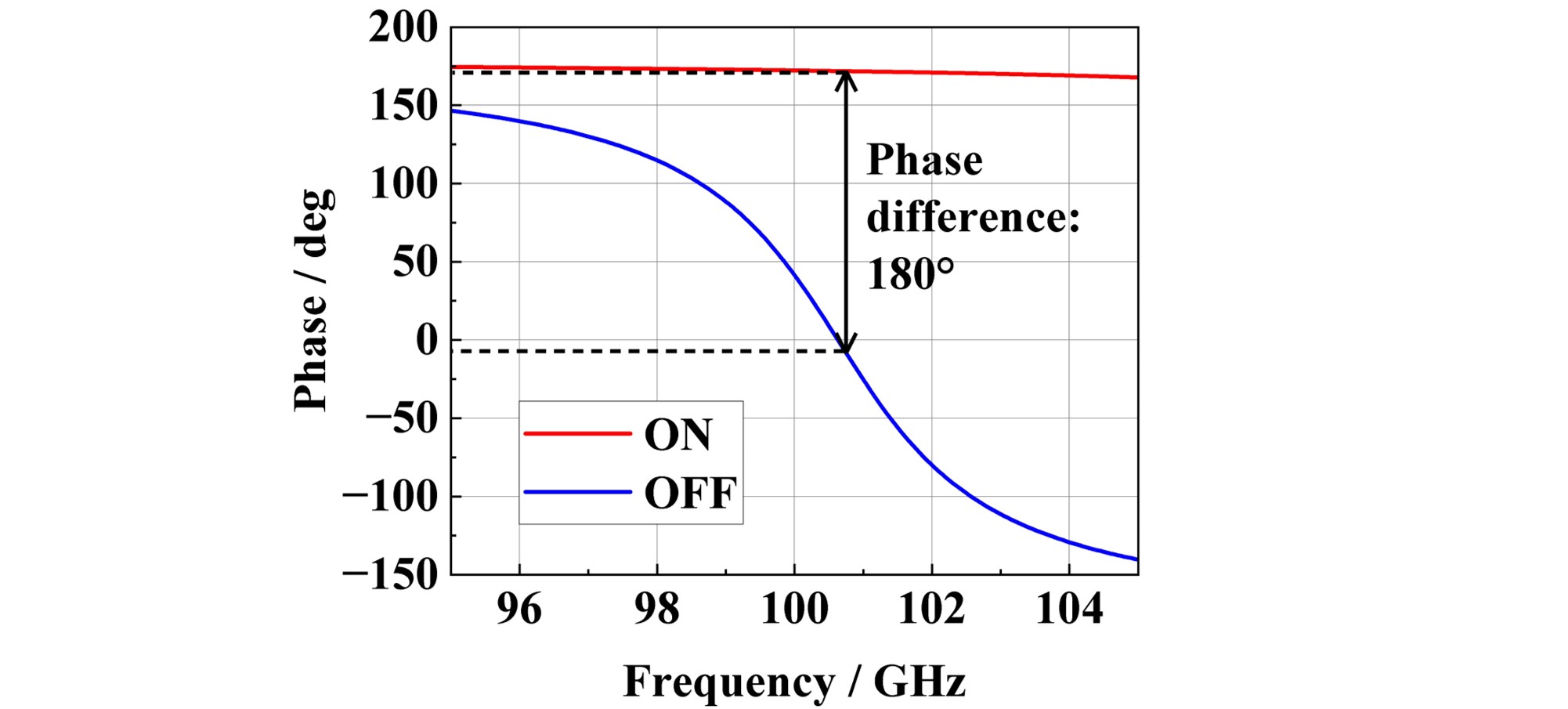}
\caption{Simulated unit cell performance at ON/OFF states: Reflection phase.}
\label{fig3a}
\end{figure}

\begin{figure}[htp]
\centering
\includegraphics[width=\columnwidth]{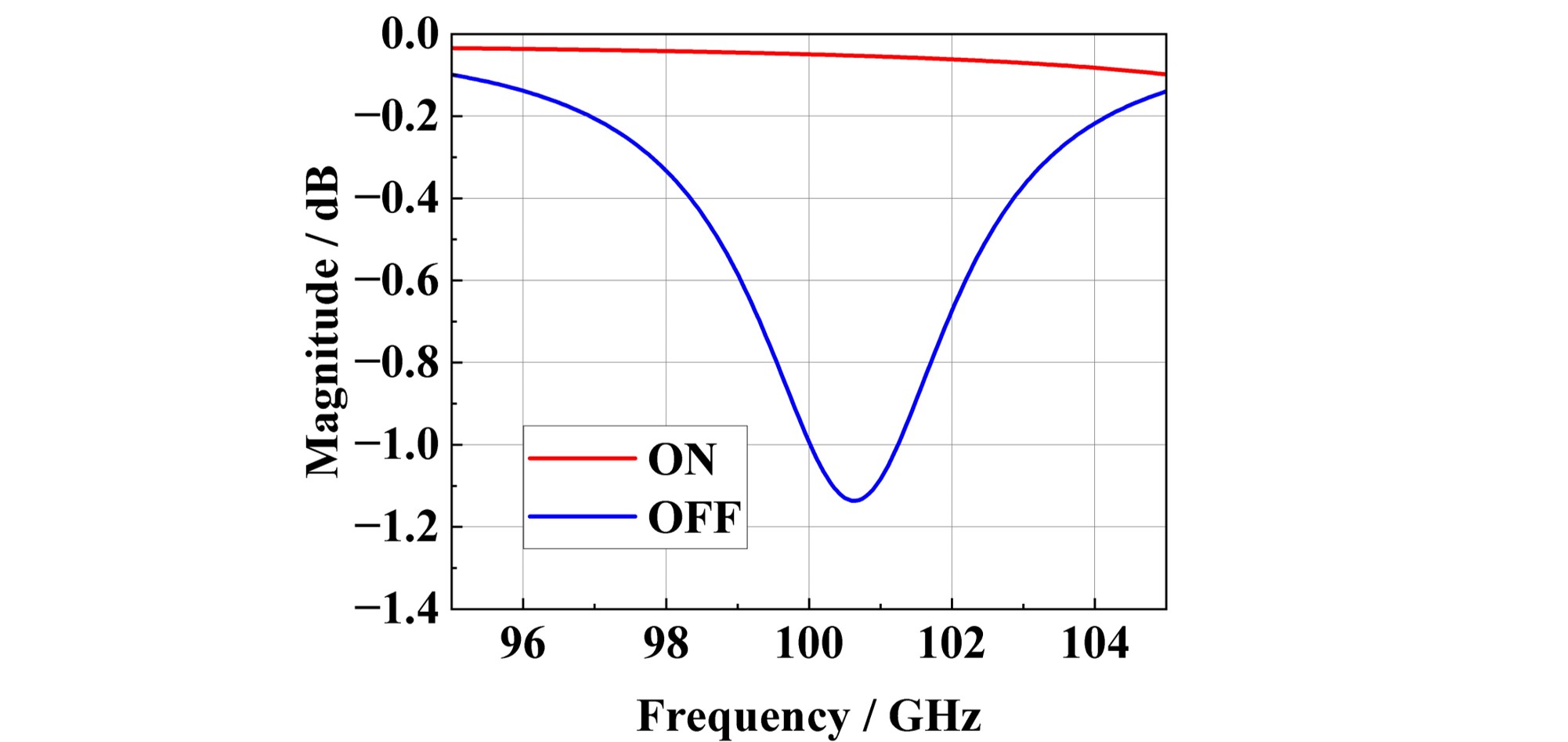}
\caption{Simulated unit cell performance at ON/OFF states: Reflection magnitude.}
\label{fig3b}
\end{figure}

The proposed unit cell was analyzed using a Floquet port simulation in HFSS to evaluate its reflection characteristics under periodic boundary conditions. 
The simulated phase response is shown in Fig.~\ref{fig3a}, while the reflection magnitude is presented in Fig.~\ref{fig3b}. 
At the operating frequency of 100.75~GHz, a phase difference of 180$^\circ$ is achieved between the two switching states, and the reflection magnitude remains better than $-1.2$~dB for both conditions.

\begin{figure}[htp]
\centering
\includegraphics[width=\columnwidth]{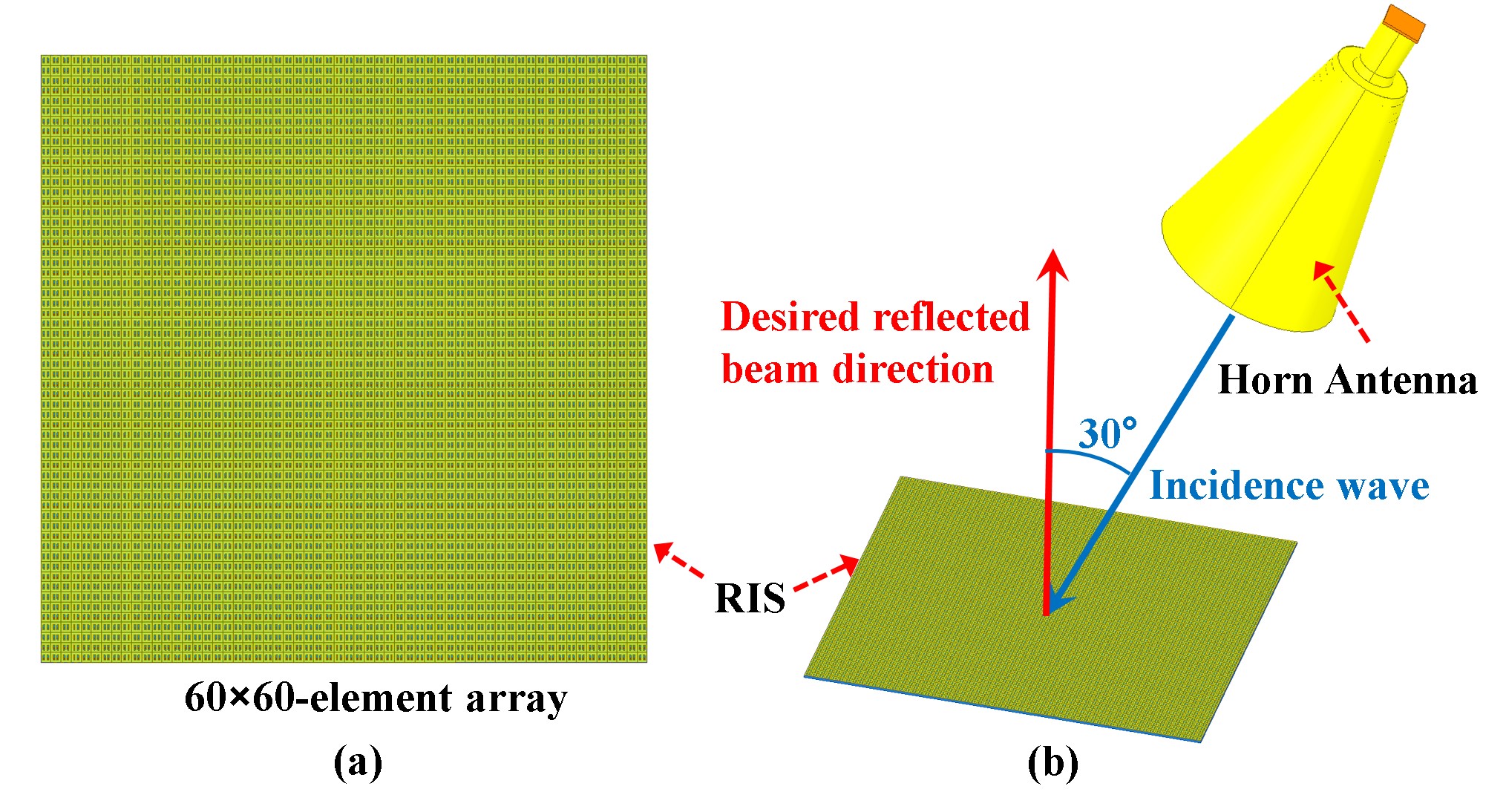}
\caption{Simulation configuration of the proposed 60×60 RIS array.}
\label{fig10}
\end{figure}

\subsection{Array Analysis}
The proposed unit cell was used to design a reconfigurable, 60×60-element array, as shown in Fig.~\ref{fig10}(a).

The phase distribution of the RIS can be theoretically calculated using the equation presented in~\cite{gros2021}:
\begin{equation}
\varphi_{ij} = k \left| \vec{r}_{ij}^{\,e} - \vec{r}^{\,f} \right| - k \left( \vec{u}_0 \cdot \vec{r}_{ij}^{\,e} \right) + \Delta\varphi
\end{equation}
where $\varphi_{ij}$ is the ideal continuous phase distribution for the unit cell located at the Cartesian coordinate $(i,j)$, 
$k$ is the wavenumber in free space, $\vec{r}_{ij}^{\,e}$ and $\vec{r}^{\,f}$ denote the element position vector and excitation source position vector, respectively, 
$\vec{u}_0$ is the unit vector of the desired reflected beam direction, and $\Delta\varphi$ is a flexible phase offset to adjust the reference phase level.

The proposed RIS unit cell exhibits two reflection states corresponding to the insulating and metallic phases of VO$_2$. Therefore, the continuous ideal phase distribution should be quantized into two discrete reflection states, $\varphi_1$ and $\varphi_2$, corresponding to the two switching conditions of the unit cell. 
The quantization process can be expressed as:
\begin{equation}
\varphi_{ij} =
\begin{cases}
\varphi_1, & \text{if } |\varphi_{ij} - \varphi_1| \leq |\varphi_{ij} - \varphi_2| \\
\varphi_2, & \text{otherwise}
\end{cases}
\end{equation}

\begin{figure}[htp]
\centering
\includegraphics[width=\columnwidth]{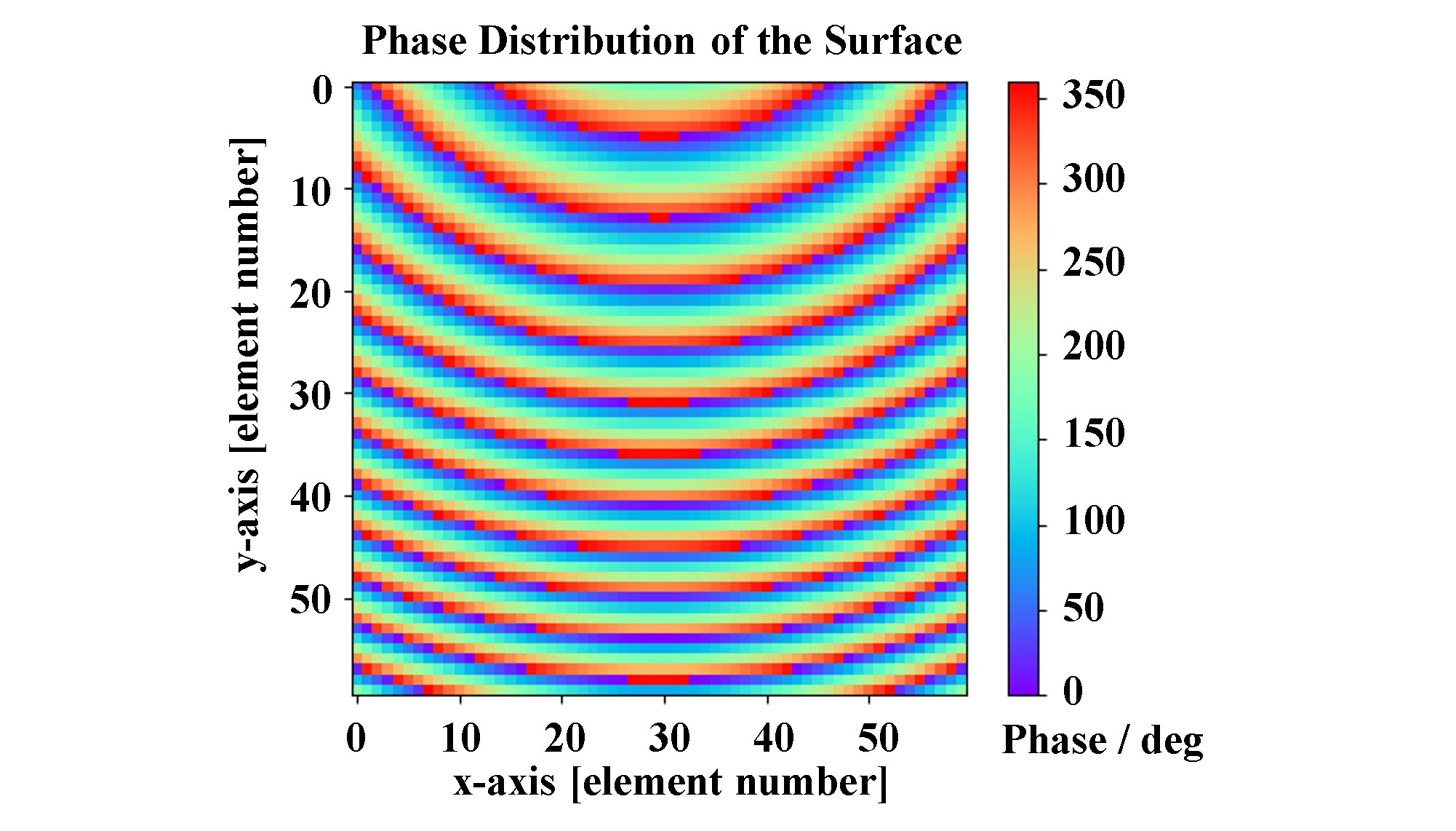}
\caption{Phase Distribution of the Surface. The horizontal and vertical axes represent the number of elements in the x and y directions, respectively. The color bar indicates the phase in degrees.}
\label{fig6}
\end{figure}

\begin{figure}[htp]
\centering
\includegraphics[width=\columnwidth]{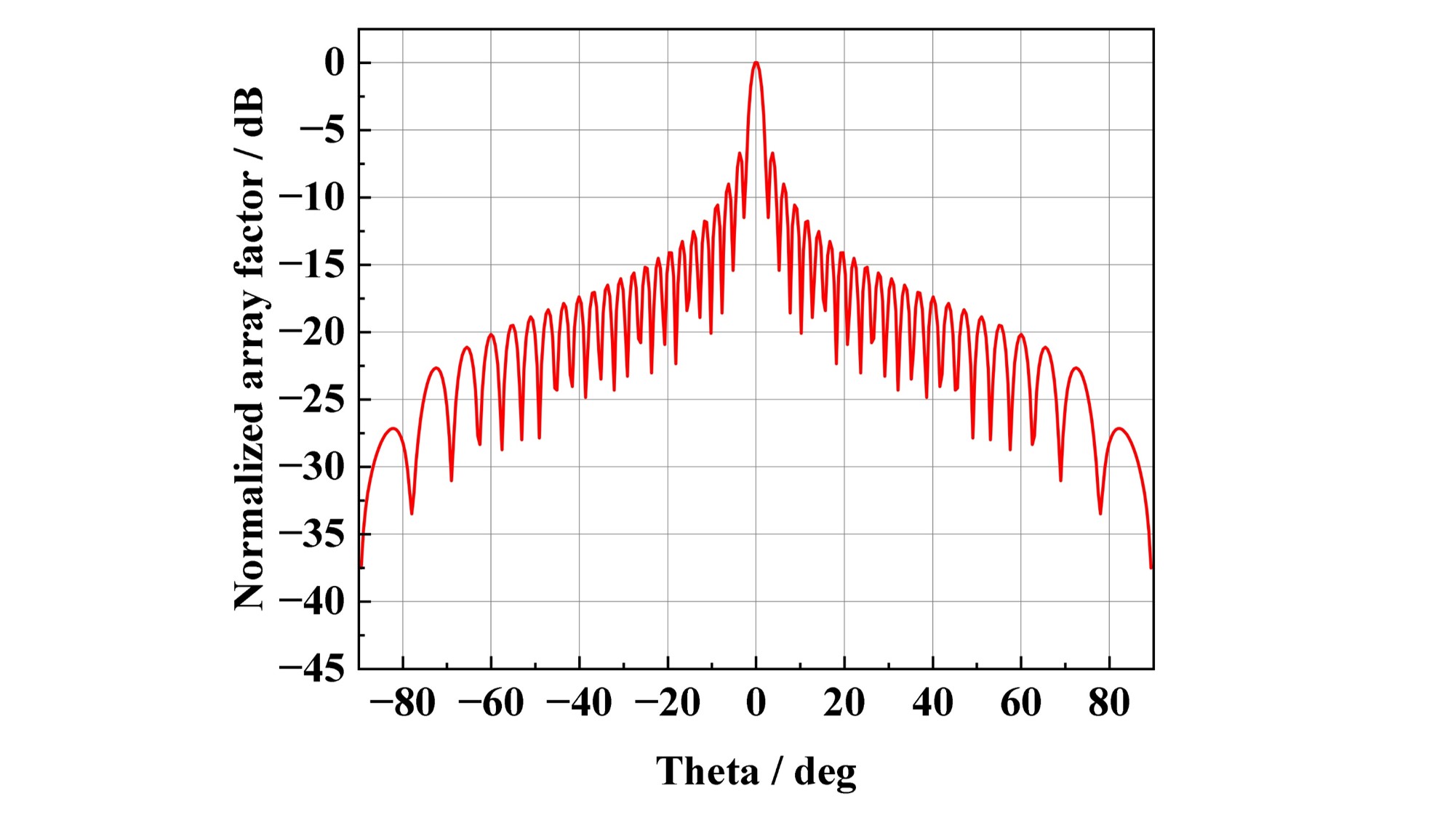}
\caption{Normalized array factor versus theta. }
\label{fig7}
\end{figure}

\begin{figure}[htp]
\centering
\includegraphics[width=\columnwidth]{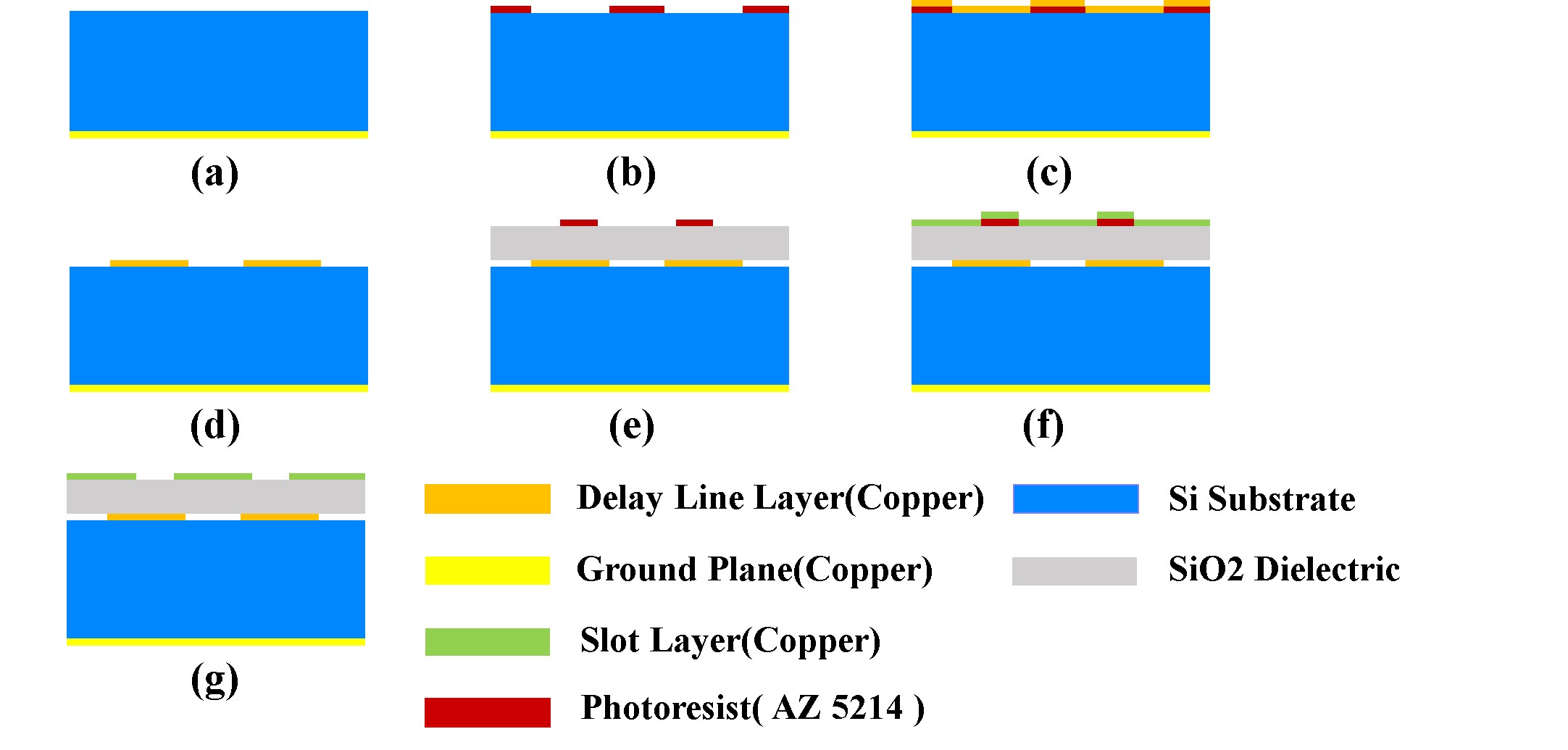}
\caption{Fabrication process of the proposed RIS. (a) Prepared 300-um-thick high-resistivity silicon wafer with one copper layer sputtered on one side. (b)Lithography. (c)Sputtering of copper. (d)Lift-off. (e)-(g) Deposited SiO2 and repeated the processes of (b)-(d).}
\label{fig8}
\end{figure}
\begin{figure}[htp]
\centering
\includegraphics[width=\columnwidth]{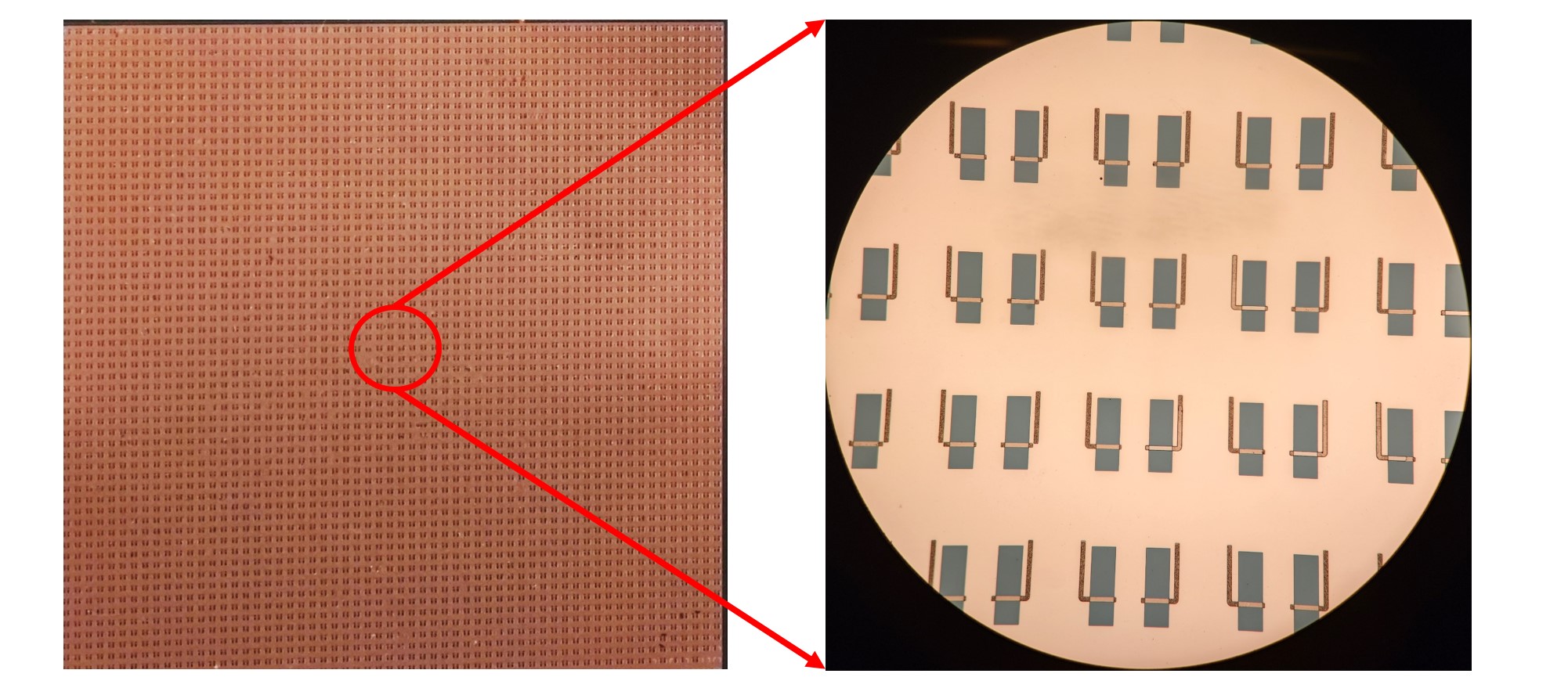}
\caption{Photograph of the fabricated RIS-on-chip array and microscopic view of the unit cells.}
\label{fig9}
\end{figure}

\begin{figure}[htp]
\centering
\includegraphics[width=\columnwidth]{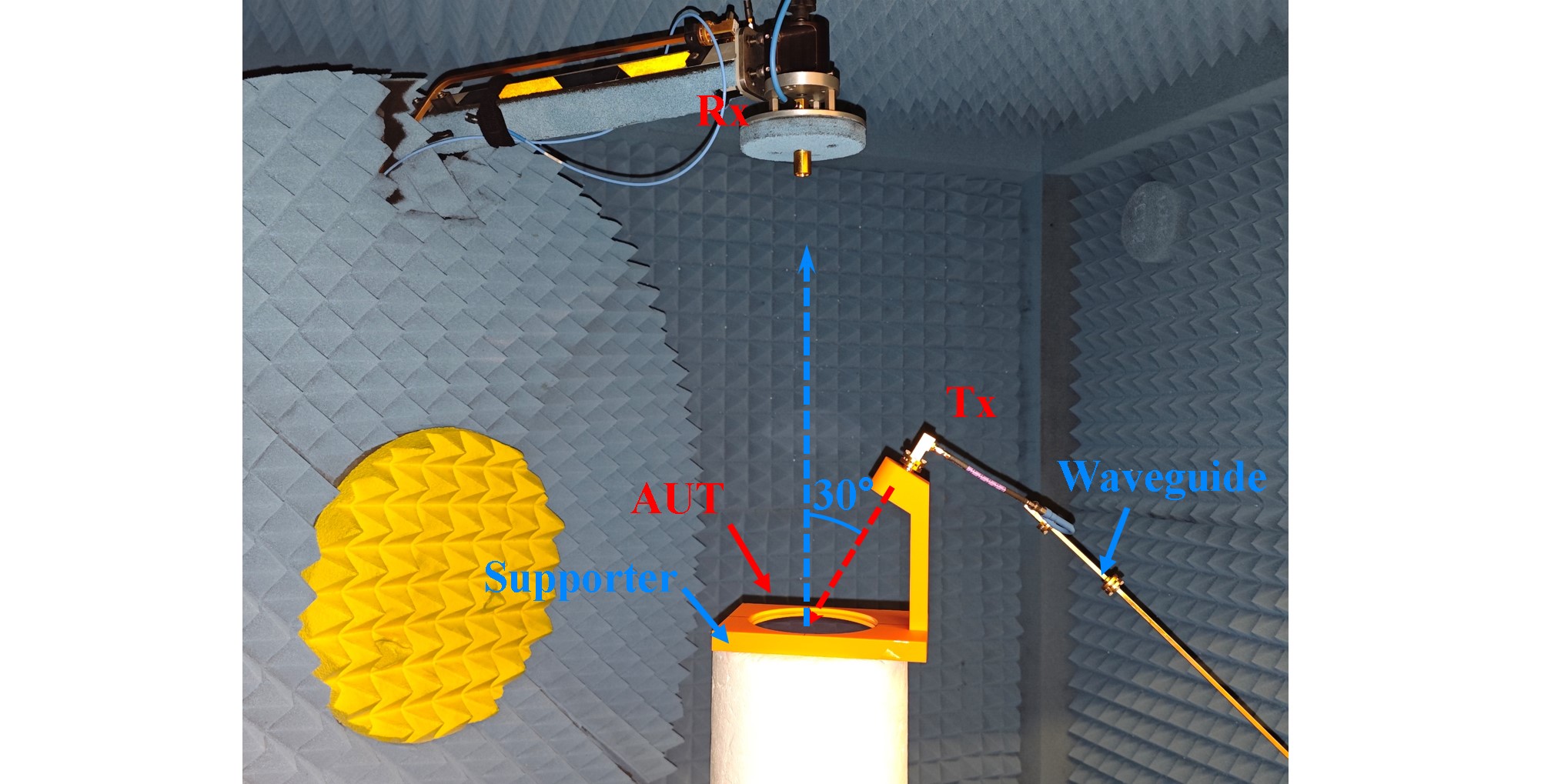}
\caption{Measurement setup in u-Lab mm-Wave anechoic chamber.}
\label{fig5}
\end{figure}
In other words, the ideal continuous phase profile is quantized into two discrete levels separated by 180$^\circ$, 
which are obtained from the full-wave simulation results of the proposed unit cell. 
The simulation setup of the RIS array is shown in Fig.~\ref{fig10}(b). 
A horn antenna (SFH-10) is used as the excitation source for the array. An incident plane wave with an angle of 30° is considered, and the reflected beam is designed to be directed at normal incidence (0°).
Based on this 1-bit quantization, the complete RIS phase distribution is constructed to form the desired reflected beam direction, as shown in Fig.~\ref{fig6}.

Fig.~\ref{fig7} shows the calculated normalized array factor, exhibiting a clear main lobe directed at 0°, which verifies the expected beam-steering performance.

\section{Fabrication and Results}
The proposed RIS-on-chip was fabricated using a CMOS-compatible process, as illustrated in Fig.~\ref{fig8}. 
First, a copper ground plane was deposited on a high-resistivity silicon substrate. 
A patterned copper layer was then added to form the delay lines, followed by PECVD deposition of a SiO$_2$ dielectric layer for insulation. 
Finally, the top slot layer was defined using photolithography, metal deposition, and lift-off, completing the multilayer stack. 
The overall process demonstrates that the proposed structure can be realized with standard thin-film and lithographic steps compatible with semiconductor technology. 

After completing the fabrication steps described above, the final RIS-on-chip array was obtained as shown in Fig.~\ref{fig9}. The left image shows the complete array, while the right image presents a microscopic view of several unit cells. 

\begin{figure}
\centering
\includegraphics[width=\columnwidth]{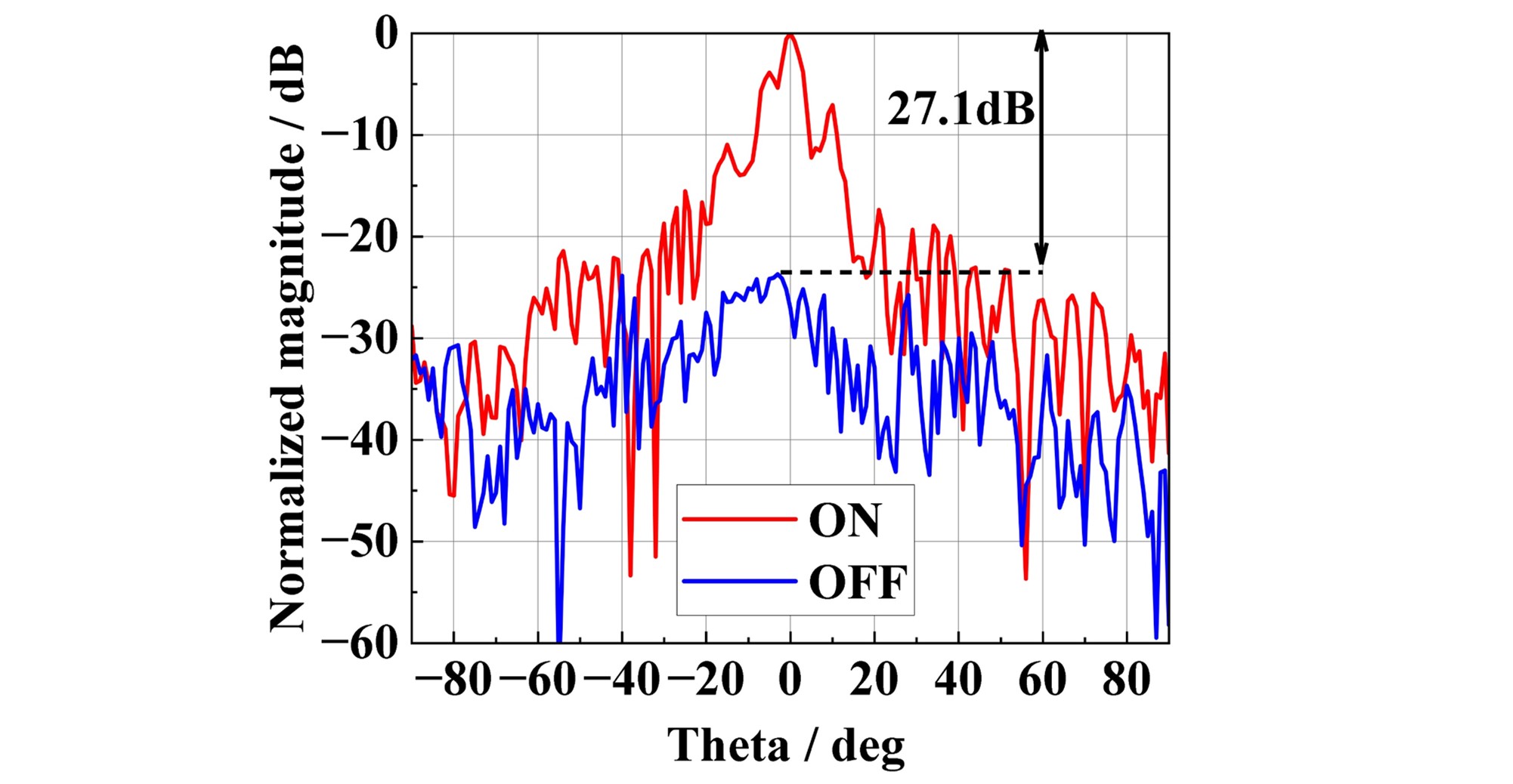}
\caption{Measured normalized reflection magnitude for the ON and OFF (PEC) states.}
\label{fig11}
\end{figure}

The measurement setup is shown in Fig.~\ref{fig5}. The distance between the transmitting horn antenna (Tx) and the RIS was calculated to be 12~cm, ensuring optimal efficiency. A custom-designed 3D-printed supporter was used to precisely maintain this separation. 
The measurement was conducted in the u-Lab millimeter-wave anechoic chamber, which supports frequencies up to 110~GHz. 
A standard horn antenna (SFH-10) was employed as the receiving antenna (Rx), which was mechanically rotated from $-90^\circ$ to $90^\circ$ to scan the elevation angle $\theta$. 
The measured normalized reflection magnitude for the ON and OFF states (PEC equivalent) is shown in Fig.~\ref{fig11}. A reflection enhancement of 27.1~dB was observed at $\theta = 0^\circ$, confirming that the fabricated RIS-on-chip prototype operates effectively as designed.

\section{Conclusion}
This paper presents a CMOS-compatible RIS-on-chip designed for 6G applications. The unit cell uses a phase delay line loaded with VO$_2$ to realize 1-bit phase control with high reflection efficiency. 
Full-wave simulations confirm a 180° phase shift and low insertion loss, while fabrication results verify the feasibility of chip-level integration using standard processes. The measurement result demonstrates a 27.1 dB enhancement between ON and OFF states.
The proposed design offers a CMOS-compatible and low-loss solution for future sub-terahertz and 6G communication.

\bibliographystyle{IEEEtran}   
\bibliography{ref1}            

\end{document}